\documentclass[%
 reprint,
superscriptaddress,
showpacs,showkeys,preprintnumbers,
 amsmath,amssymb,
 aps,
]{revtex4-1}

\usepackage{graphicx}% Include figure files
\usepackage{dcolumn}% Align table columns on decimal point
\usepackage{bm}% bold math
\usepackage{here}
\usepackage[dvips]{color}
\usepackage{multirow}
\usepackage{ulem}
\newcommand{\abs}[1]{\left\vert{#1}\right\vert}

\makeatletter
  \def\erf{\mathop{\operator@font erf}\nolimits}
  \def\erfc{\mathop{\operator@font erfc}\nolimits}
  \def\Erf{\mathop{\operator@font Erf}\nolimits}
  \def\Shi{\mathop{\operator@font Shi}\nolimits}
  \def\Chi{\mathop{\operator@font Chi}\nolimits}
  \def\Ei{\mathop{\operator@font Ei}\nolimits}
  \def\cosec{\mathop{\operator@font cosec}\nolimits}
  \def\sech{\mathop{\operator@font sech}\nolimits}
  \def\cosech{\mathop{\operator@font cosech}\nolimits}
  \newcommand\hypgeo[2]{{}_{#1}{\operator@font F}_{#2}}
  \def\Re{\mathop{\operator@font Re}\nolimits}
  \def\Im{\mathop{\operator@font Im}\nolimits}
\makeatother
\begin{document}
%\linenumbers

%\preprint{APS/123-QED}

%======================================================
%	title
%======================================================
%\title{Spin-dependent cross section in the neutron induced $p$-wave resonance of $^{139}$La with polarized nuclear target and polarized epi-thermal neutrons}
%\title{Spin-dependent cross section in neutron induced $p$-wave resonances in La
%\title{Spin-dependent cross section in the $p$-wave resonance of $\protect\overrightarrow{\rm{^{139}La}}+\protect\overrightarrow n$}
%\title{Spin-spin interaction in the $p$-wave resonance of $\bm{{{^{139}\protect\overrightarrow{\rm{La}}}}+\protect\overrightarrow n}$}
\title{Spin dependence in the $p$-wave resonance of $\bm{{{^{139}\protect\overrightarrow{\rm{La}}}}+\protect\overrightarrow n}$}
% Force line breaks with \\

%======================================================
%	authors
%======================================================
\def\affNagoya{Nagoya University, Furocho, Chikusa, Nagoya 464-8602, Japan}
\def\affKyushu{Kyushu University, 744 Motooka, Nishi, Fukuoka 819-0395, Japan}
\def\affJAEA{Japan Atomic Energy Agency, 2-4 Shirakata, Tokai, Ibaraki 319-1195, Japan}
\def\affTokyoTech{Tokyo Institute of Technology, Meguro, Tokyo 152-8551, Japan}
\def\affIbaraki{Ibaraki University, 2-1-1 Bunkyo, Mito, Ibaraki 310-8512. Japan}
\def\affRCNP{Osaka University, Ibaraki, Osaka 567-0047, Japan}
\def\affIndiana{Indiana University, Bloomington, Indiana 47401, USA}
\def\affLosAlamos{Los Alamos National Laboratory, Los Alamos, NM 87545, USA}
\def\affKEK{High Energy Accelerator Research Organization,1-1 Oho, Tsukuba, Ibaraki 305-0801, Japan}
\def\affTohoku{Tohoku University, 2-1-1 Katahira, Aoba-ku, Sendai, 980-8576 Japan}
\def\affCross{Comprehensive Research Organization for Science and Society, Tokai, Ibaraki 319-1106, Japan}
\def\affSouthCarolina{University of South Carolina, Columbia, South Carolina 29208, USA}

\author{T.~Okudaira}
%\email{okudaira@phi.phys.nagoya-u.ac.jp}
\affiliation{\affNagoya}
\affiliation{\affJAEA}

\author{R.~Nakabe}
%\email{okudaira@phi.phys.nagoya-u.ac.jp}
\affiliation{\affNagoya}

%\author{C.~Auton}
%\email{okudaira@phi.phys.nagoya-u.ac.jp}
%\affiliation{\affIndiana}
%\affiliation{\affNagoya}

\author{S.~Endo}
\affiliation{\affNagoya}
\affiliation{\affJAEA}

\author{H.~Fujioka}
\affiliation{\affTokyoTech}

\author{V. Gudkov}
\affiliation{\affSouthCarolina}

\author{I.~Ide}
\affiliation{\affNagoya}

\author{T.~Ino}
\affiliation{\affKEK}

\author{M.~Ishikado}
\affiliation{\affCross}

\author{W.~Kambara}
\affiliation{\affJAEA}

\author{S.~Kawamura}
\affiliation{\affNagoya}
\affiliation{\affJAEA}

\author{R.~Kobayashi}
\affiliation{\affIbaraki}

\author{M.~Kitaguchi}
\affiliation{\affNagoya}

\author{T.~Okamura}
\affiliation{\affKEK}

\author{T.~Oku}
\affiliation{\affJAEA}
\affiliation{\affIbaraki}

\author{J.~G.~Otero~Munoz}
\affiliation{\affIndiana}

\author{J. D.~Parker}
\affiliation{\affCross}

\author{K.~Sakai}
\affiliation{\affJAEA}

\author{T.~Shima}
\affiliation{\affRCNP}

\author{H.~M.~Shimizu}
\affiliation{\affNagoya}

\author{T.~Shinohara}
\affiliation{\affJAEA}

\author{W.~M.~Snow}
\affiliation{\affIndiana}

\author{S.~Takada}
\affiliation{\affTohoku}
\affiliation{\affJAEA}

\author{Y.~Tsuchikawa}
\affiliation{\affJAEA}

\author{R.~Takahashi}
\affiliation{\affJAEA}

\author{S.~Takahashi}
\affiliation{\affIbaraki}
\affiliation{\affJAEA}

\author{H.~Yoshikawa}
\affiliation{\affRCNP}

\author{T.~Yoshioka}
%\email{yoshioka@epp.phys.kyushu-u.ac.jp}
\affiliation{\affKyushu}

%======================================================

\date{\today}% It is always \today, today,

%======================================================
%	abstract
%======================================================

\begin{abstract}
We measured the spin dependence in a neutron-induced $p$-wave resonance by using a polarized epithermal neutron beam and a polarized nuclear target.  Our study focuses on the 0.75~eV $p$-wave resonance state of $^{139}$La+$n$, where largely enhanced parity violation has been observed. We determined the partial neutron width of the $p$-wave resonance by measuring the spin dependence of the neutron absorption cross section between polarized $^{139}\rm{La}$ and polarized neutrons. Our findings serve as a foundation for the quantitative study of the enhancement effect of the discrete symmetry violations caused by mixing between partial amplitudes in the compound nuclei.

%A very large enhanced parity violation was reported at the 0.75~eV $p$-wave resonance formed after the neutron absorption reaction of $^{139}$La due to the mixing of the partial waves. We observed a spin dependent cross section between polarized $^{139}$La and polarized epi-thermal neutrons using a pulsed neutron beam, especially for the 0.75~eV $p$-wave resonance. The partial neutron width of the $p$-wave resonance were determined. The experimental result will provide the basis for the quantitate study of the enhancement mechanism of the discrete symmetry violation and search for the time reversal violation in the nucleon-nucleon interaction.
\end{abstract}

%======================================================
%	PACs
%======================================================
                             
\keywords{neutron induced compound nuclei,
polarized epithermal neutrons, 
polarized nuclear target}%Use showkeys class option if keyword
\maketitle

%\tableofcontents

%======================================================
%	main body
%======================================================
%%%%%%%%%%%%%%%%%%%%%%%%%%%%%%%%%%%%%%%%%%%%%%%%%%%%%%%%%%%%%%%%%%%%%%%%%%%%%%%%
%	Introduction
%%%%%%%%%%%%%%%%%%%%%%%%%%%%%%%%%%%%%%%%%%%%%%%%%%%%%%%%%%%%%%%%%%%%%%%%%%%%%%%%
\section{Introduction}
The spin dependence of the strong interaction between a neutron and a nucleus can lead to a spin-dependent cross section proportional to $\bm{\sigma} \cdot \bm{I}$, where $\bm{\sigma}$ and $\bm{I}$ are unit vectors parallel to the spins of the neutron and the nucleus, respectively. This spin-dependent cross section can be observed through a spin-dependent transmission through a polarized nuclear target. At a neutron-nucleus resonance, this observable can directly determine the spin of compound resonance states. Consequently, it has been employed in measuring $s$-wave resonances for a select few nuclides, utilizing both a polarized neutron beam and a polarized target~\cite{Keyworth1973U, Keyworth1973Np, Alfimenkov1993}. In the case of $p$-wave resonances, the spin-dependent cross section imparts valuable information not only regarding the spin of the resonance but also the partial neutron widths. \\
These widths enable the exploration of symmetry violation enhancement effects in the compound nucleus. Enhancements in parity violation, exceeding magnitudes of nucleon-nucleon interactions by a factor of 10$^6$, have been observed in $p$-wave resonances of nuclei with a mass number of medium-heavy or heavy nuclei~\cite{Mitchell2001}. These enhancements can be understood as a result of parity mixing between $s$- and $p$- compound nuclear resonances. This is referred to as  the $s$-$p$ mixing model~\cite{Mitchell2001, sus82E, BUNAKOV198393}. Theory suggests that this mechanism can lead to an enhancement of fundamental time reversal violation, which could be utilized to search for beyond the Standard Model physics by measuring a T-odd cross section at the $p$-wave resonance using a polarized target and a polarized neutron beam~\cite{Gudkov1991}. We can quantify the enhancement factors associated with both P- and T-violations by determining the partial neutron width~\cite{gud92, gud17,okuda18, okuda21, yama20, koga22, endo22, okuda2023}.\\
This paper presents the first measurement of the spin-dependent cross section at the $p$-wave compound resonance, employing a polarized epithermal neutron beam and a polarized nuclear target. As our target nucleus, we selected $^{139}$La, which displays an exceedingly large enhanced parity violation at the 0.75~eV $p$-wave resonance~\cite{LANL91}.

\section{Experiment}
%%%%%%%%%%%%%%%%%%%%%%%%%%%%%%%%%%%%%%%%%%%%%%%%%%%%%%%%%%%%%%%%%%%%%%%%%%%%%%%%
%	Experiment : Experimental Setup
%%%%%%%%%%%%%%%%%%%%%%%%%%%%%%%%%%%%%%%%%%%%%%%%%%%%%%%%%%%%%%%%%%%%%%%%%%%%%%%%
\subsection{Experimental setup}
\begin{figure}[htbp]
	\centering
	\includegraphics[width=0.95\linewidth]{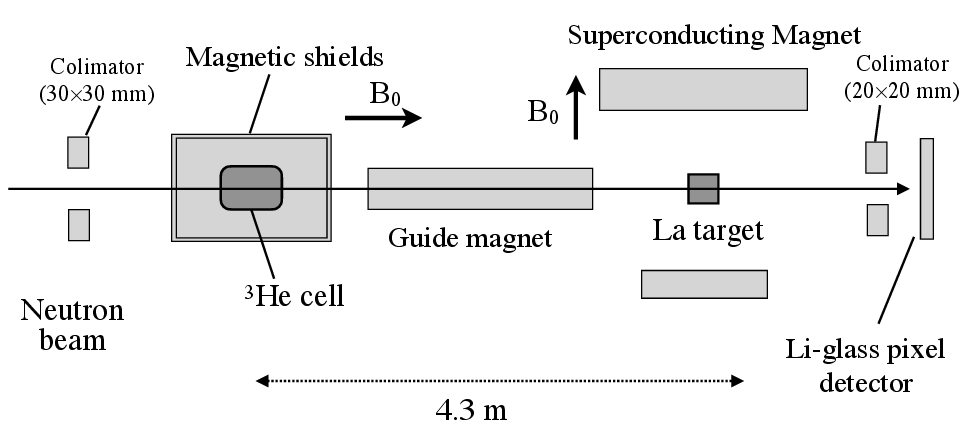}
	\caption[]{
	Experimental setup.  
	}
	\label{setup}
\end{figure}

The experiment was performed with a pulsed epi-thermal neutron beam at the RADEN beamline of the Material and Life Science Experimental Facility (MLF) at the Japan Proton Accelerator Research Complex (J-PARC)~\cite{Shinohara2020}.   %In principle, the nuclear polarization can be obtained from the spin dependent cross section of $s$-wave components by using the resonance parameters of $^{139}$La+n states and the spin dependence of the potential scattering radius. However, due to the lack of sufficient data for the spin dependence of the potential scattering radius and resonance parameters, the nuclear polarization estimated from the temperature is used in this study. 
The experimental setup is depicted in Fig.~\ref{setup}. The La target is placed 23.0~m from the moderator surface. A 2.0 cm cubic lanthanum metal cooled with a dilution refrigerator was used as the target. A 6.8 Tesla transverse magnetic field was applied using a superconducting magnet to polarize the target nuclei. The neutron beam, collimated to a 3 cm by 3 cm size, was stripped of thermal neutrons using a cadmium filter upstream of the beamline to reduce the heat load on the La target induced by the neutron beam. The beam was polarized with a neutron polarizer using polarized $^3$He gas ($^3$He spin filter), located 4.3 m upstream of the polarized target. The $^3$He spin filter was polarized using the spin exchange optical pumping method (SEOP) with a 110 W laser system constructed outside of the beamline and then installed on the beamline with a coil and a double magnetic shield to maintain the $^3$He polarization~\cite{He3Okudaira}. The $^3$He cell was  45~mm in diameter by 70~mm in length and the pressure was 0.31~MPa. The neutron beam, longitudinally polarized by the $^3$He spin filter, was guided using a guide magnet. The spin direction was adiabatically rotated to the transverse direction utilizing the stray magnetic field of the superconducting magnet. The neutron spin was flipped every 30~minutes by flipping the spin of $^3$He gas using adiabatic fast passage (AFP) NMR. The loss of the $^3$He polarization was $4 \times 10^{-5}$ per flip, which was negligibly small. Transmitted neutrons were recorded in list mode using a 256-pixel lithium glass scintillator detector located at 24.71~m from the moderator surface~\cite{Liglass2015}. Downstream of the La target, another collimator was installed to reduce the beam divergence. The neutron energy $E_{n}$ was determined using the neutron time-of-flight (TOF) and the flight path length. The proton beam power was 750~kW during the experiment.
 \par
\begin{figure}[htbp]
	\centering
	\includegraphics[width=0.70\linewidth]{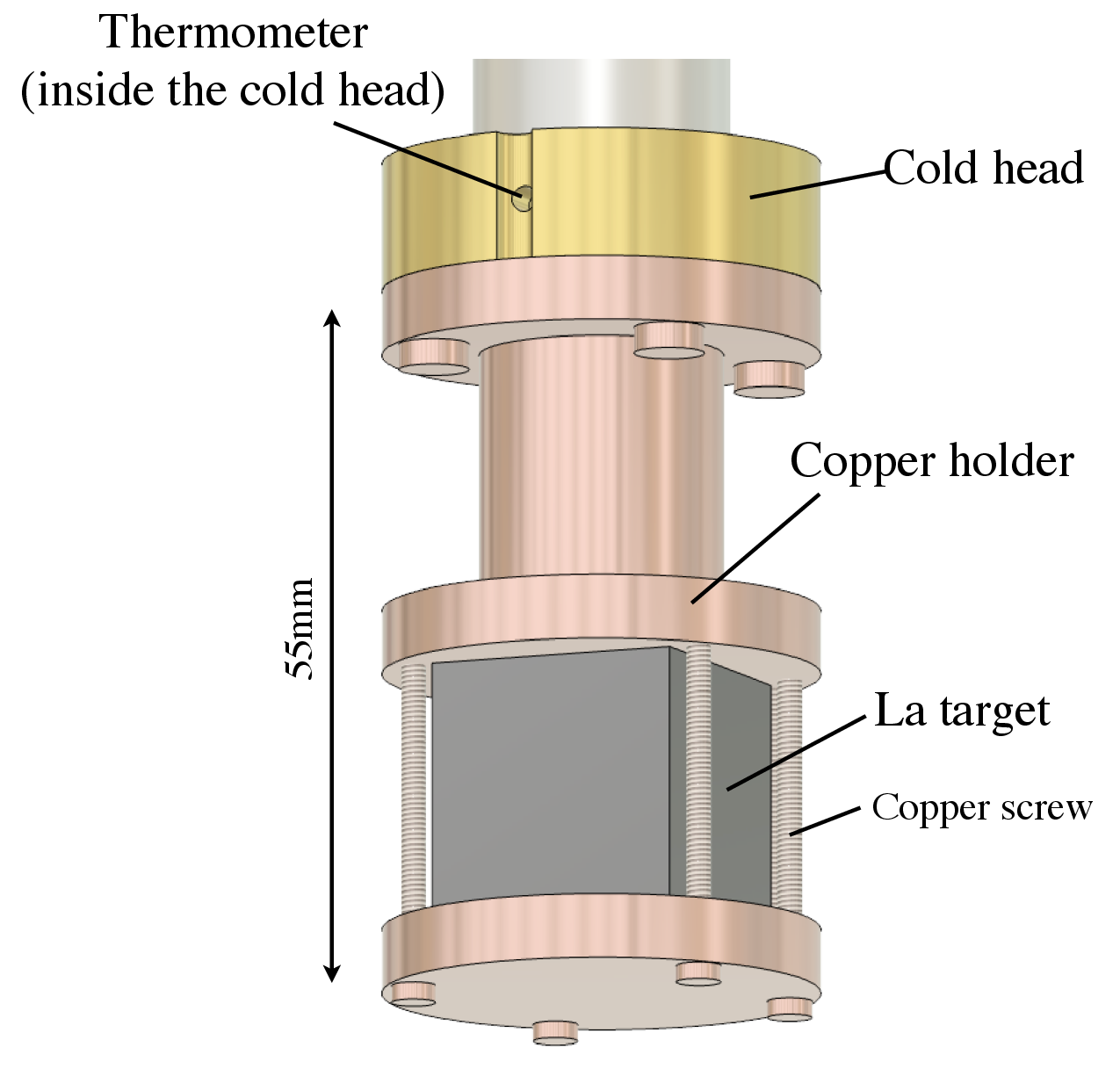}
	\caption[]{
	Configuration around the La target. The thermometer was installed in the cold head.
	}
	\label{target}
\end{figure}

 Figure~\ref{target} illustrates the configuration around the La target. The La target was held in place between upper and lower copper holders, fastened using copper screws. The upper holder was connected to the cold head of the dilution refrigerator, enabling cooling of the La target through thermal conduction. The temperature $T$ was monitored with a ruthenium oxide thermometer installed in the cold head. We performed the experiment in two conditions: (a) low temperature condition ($T=67~\rm{mK}$ ) and (b) high temperature condition ($T=1~\rm{K}$). The measurement times were 22~hours and 6~hours, respectively. In the condition (a), the temperature increase by the beam irradiation to the La target was approximately 1~mK, indicating that the temperature difference between the cold head and the La target can be considered negligible. The temperature fluctuation was also around 1~mK, which was caused by beam interruptions in the accelerator due to malfunctions.

%Consequently, this paper uses a $^{139}$La nuclear polarization of 4.4\%, calculated assuming a Boltzmann distribution \comment{with its nuclear spin and magnetic moment listed in Table~\ref{nucpara}} in the temperature measured at the cold head and the applied magnetic field. The fluctuation of the nuclear polarization derived from the temperature fluctuation of 1~mK is estimated as 0.07\%. As will be discussed later, the measurement result of the spin dependent cross section supports the nuclear polarization estimated based on the target temperature.}


%%%%%%%%%%%%%%%%%%%%%%%%%%%%%%%%%%%%%%%%%%%%%%%%%%%%%%%%%%%%%%%%%%%%%%%%%%%%%%%%
%	Experiment : Measurement
%%%%%%%%%%%%%%%%%%%%%%%%%%%%%%%%%%%%%%%%%%%%%%%%%%%%%%%%%%%%%%%%%%%%%%%%%%%%%%%%
\subsection{Measurement of the asymmetry}
The cross sections for parallel and antiparallel polarized neutron and nucleus can be written with the spin-independent cross section $\sigma_0$ and spin-dependent cross section $\sigma_{\rm S}$ as 
\begin{eqnarray}
%\sigma_{\pm}=\frac{4\pi}{k}({\rm{Im}}A\pm {\rm{Im}}B),
\sigma_{\pm}=\sigma_0 \pm \sigma_{\rm{S}},
\label{spindependence}
\end{eqnarray}
where ${+}$ and ${-}$ denote the parallel and antiparallel spins, respectively. The asymmetry of neutron counts for parallel and anti-parallel spins transmitted through the polarized lanthanum target, defined as
\begin{eqnarray}
\varepsilon_{\rm{S}}=\frac{N_--N_+}{N_-+N_+},
\label{ratio}
\end{eqnarray}
where $N_-$ and $N_+$ are the neutron counts for parallel and anti-parallel spins, was measured. The neutron counts $N_{\pm}$ are expressed using the neutron polarization $P_n$ and nuclear vector polarization $P_I$ as 
\begin{eqnarray}
N_{\pm}=\frac{1\pm P_n}{2}N \epsilon \exp\left( (\sigma_0 \pm P_I\sigma_{\rm{S}})\rho d\right),
\label{counts}
\end{eqnarray}
 where $N$, $\epsilon$, $\rho$, and $d$ are number of incident neutrons, detection efficiency of a neutron detector, number density of the nuclear target, and the thickness of the nuclear target, respectively. % The asymmetry $A_{\rm s}$ can be described using Eq.{\ref{counts}} in the case of $P_I \sigma_{\rm s} \rho d<<1 $ as  
The spin-dependent asymmetry $\varepsilon_{\rm S}$ can be described using Eq.{\ref{counts}} as
\begin{eqnarray}
%A_{\rm{s}}=P_n\tanh \left( P_I \sigma'_{\rm S} \rho d \right) \sim P_nP_I \sigma'_{\rm S} \rho d.
\varepsilon_{\rm{S}}=P_n\tanh \left( P_I \sigma_{\rm S} \rho d \right).
\label{EqAs}
\end{eqnarray}
%Here, $P_I \rho d=2.3\times10^{-3}$/barn, and this approximation can be adapted unless $\sigma_{0,{\rm S}}$ is very much larger than the spin independent cross section (20~barn for the thermal neutron).  

In this paper, the measurement and analysis were performed using resonance parameters of La+n reactions listed in Table~\ref{resopara}, which were recently measured by Endo {\it{et al}.}~\cite{EndoReso2023} using both neutron transmission and ($n$, $\gamma$) reaction with an intense pulsed neutron beam at J-PARC.\\\\

\begin{figure}[htbp]
	\centering
	\includegraphics[width=0.95\linewidth]{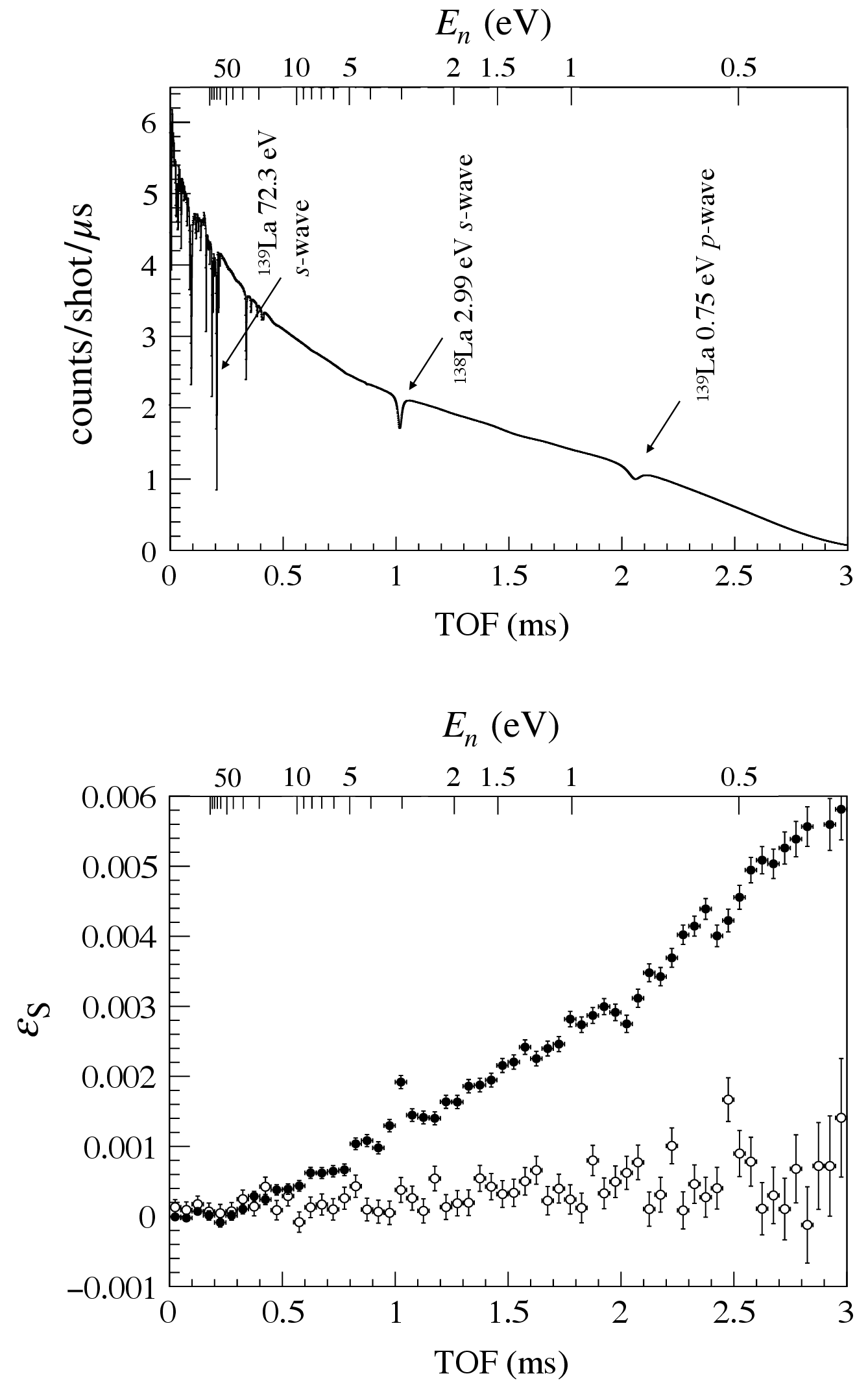}
	\caption[]{
	TOF spectra of the transmitted neutrons (top figure) and the spin-dependent asymmetries (bottom figure). Black and white points denote the asymmetries in the conditions (a) and (b), respectively, in the bottom figure.  
	}
	\label{TOF_Asy}
\end{figure}

\begin{table*}[htbp]
\begin{center}
	\begin{tabular}{|c||c|c|c|c|c|}
	%\multirow{2}{*}{$r$} \\ 
	%\multicolumn{6}{c||}{published values} \\
	\cline{1-6}
	Isotope
	&
	$E_0\,[{\rm eV}]$ & 
	$J$ & 
	$l$ & 
	$\Gamma^{\gamma}\,[{\rm meV}]$ &
	$g\Gamma^{n}\,[{\rm meV}]$\\
	\hline
	$\rm{^{139}La}$ & $-38.8\pm 0.4$ & $4$ & $0$ &$60.3\pm0.5$ & 346$\pm$10\\
	$\rm{^{139}La}$ & $0.750\pm0.001$ &4& $1$ & $41.6\pm0.9$ &  $(3.67 \pm0.05 )\times 10^{-5}$ \\
	$\rm{^{138}La}$ & $2.99\pm0.01$ &11/2 & $0$ & $95\pm6$ &$0.65\pm0.03$ \\
	$\rm{^{139}La}$ & $72.30\pm0.01$ &3 & $0$ & $64.1\pm3.0$ &$13.1\pm0.7$ \\
	\hline
	\end{tabular}
	\caption{
	The resonance parameters of La+$n$ for low energy neutrons. The resonance parameters $E$, $J$, $l$, $\Gamma^\gamma$, $g$, and $\Gamma^n$ are resonance energy, resonance spin, orbital angular momentum, $\gamma$-width, {\it{g}}-factor, and neutron width, respectively. The parameters of $^{138}$La and $^{139}$La are taken from Ref.~\cite{mughabghab} and Ref.~\cite{EndoReso2023}, respectively.
	}
	
	\label{resopara}
\end{center}	
\end{table*}
Figure~\ref{TOF_Asy} shows the TOF spectra of the transmitted neutrons and asymmetry $\varepsilon_{\rm S}$ in conditions (a) and (b).  We observed a significant asymmetry in condition (a), corresponding to a high nuclear polarization, while the asymmetry disappeared in condition (b) due to the lower nuclear polarization. The peak and dip structures were observed at the 2.99~eV and 0.75~eV resonances. The global structure observed less than 0.3~ms is attributed to the spin-dependent cross section of the negative $s$-wave resonance.\\

\subsection{Neutron polarization}
\begin{figure}[htbp]
	\centering
	\includegraphics[width=0.9\linewidth]{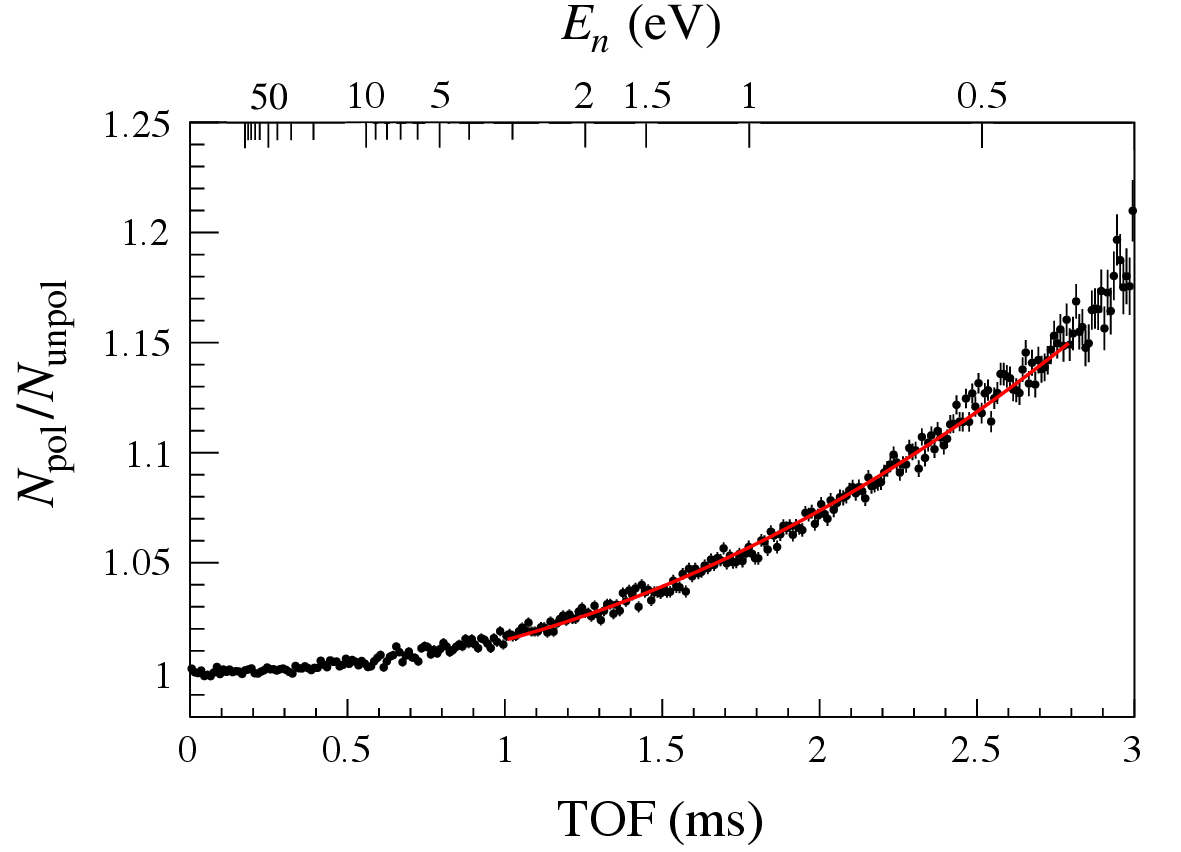}
	\caption[]{
	Ratio of the counts of transmitted neutrons for the polarized to the unpolarized $^3$He spin filter.  The curved line shows the best fit.
	}
	\label{Hetrans}
\end{figure}

\begin{figure}[htbp]
	\centering
	\includegraphics[width=0.9\linewidth]{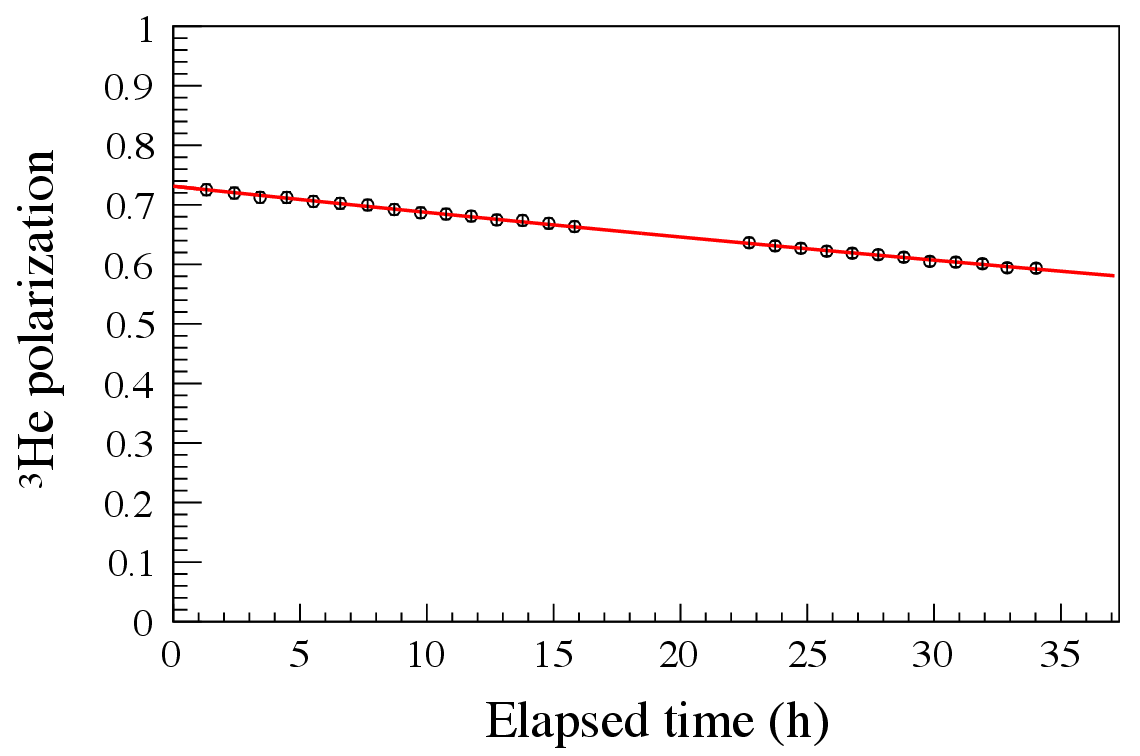}
	\caption[]{
	$^3$He polarization versus elapsed time from the beginning of the measurement.  The curved line shows the fit result by an exponential function. The measurement was not conducted from 16~h to 22~h due to a liquid He transfer for the superconducting magnet. 
	}
	\label{HeRelax}
\end{figure}

\begin{figure}[htbp]
	\centering
	\includegraphics[width=0.9\linewidth]{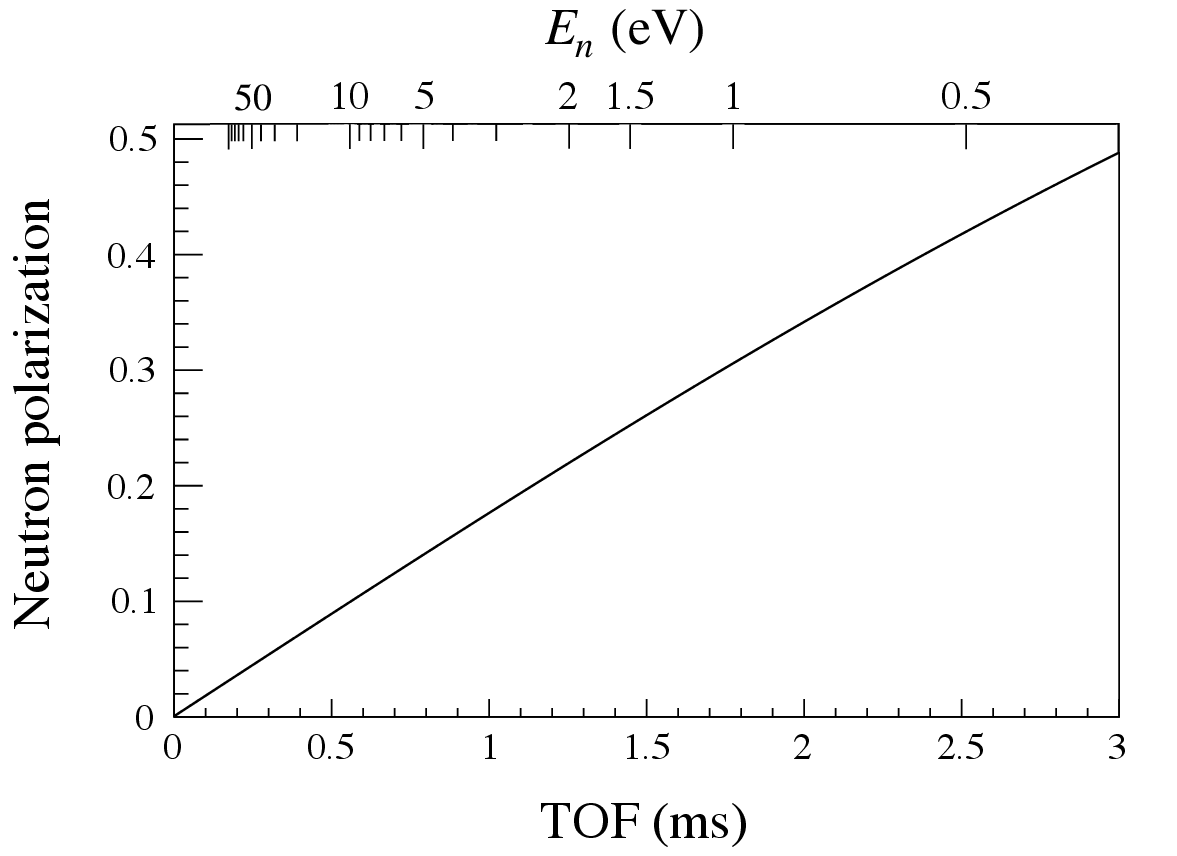}
	\caption[]{
	Neutron polarization obtained from the averaged $^3$He polarization.  
	}
	\label{Pn}
\end{figure}

The neutron polarization was obtained using the $^3$He polarization of the $^3$He spin filter. The $^3$He polarization was determined with the ratio of the transmitted neutrons for polarized and unpolarized $^3$He spin filter. The ratio of the transmitted neutrons  is described as
\begin{eqnarray}
\frac{N_{\rm{pol}}}{N_{\rm{unpol}}}=\cosh(P_{\rm{He}}(t)\rho_{\rm{He}} d_{\rm{He}} \sigma_{\rm{He}}),
\label{3Heratio}
\end{eqnarray}
where, $P_{\rm{He}}$, $\sigma_{\rm{He}}$, and $\rho_{\rm{He}} d_{\rm{He}}$ are the $^3$He polarization, neutron absorption cross section of $^3$He, and areal density of $^3$He gas, respectively.  Here, $N_{\rm{pol}}$ is defined as $N_{+}+N_{-}$ for cancelling the spin-dependent asymmetry derived from the polarization of the La target. The areal density $\rho_{\rm{He}} d_{\rm{He}}$ was obtained from the measurement of the ratio of transmitted neutrons for unpolarized $^3$He spin filter and empty glass cell as 21.4~atm$\cdot$cm. The $^3$He polarization was obtained for each flip by fitting the TOF dependence of $N_{\rm{pol}}/N_{\rm{unpol}}$ using Eq.~\ref{3Heratio} with a fit parameter of $P_{\rm{He}}$ as shown in Fig~\ref{Hetrans}. Figure \ref{HeRelax} shows the time dependence of the $^3$He polarization. The relaxation time of the $^3$He polarization $\tau$, which was obtained by fitting with $P_{\rm{He}}(t)=P_{\rm{He}}(0)\exp(-t/\tau)$, was 161~h. The averaged $^3$He polarization $\bar{P}_{\rm{He}}$ during the measurement was $(68\pm1)$\%. \\
The neutron polarization $P_n$ transmitted through the $^3$He spin filter is determined as 
\begin{eqnarray}
P_n(t)=-\tanh(P_{\rm{He}}(t)\rho_{\rm{He}} d_{\rm{He}} \sigma_{\rm{He}}).
\label{ratio}
\end{eqnarray}
Figure~\ref{Pn} shows an averaged neutron polarization $\bar{P}_{n}$ as a function of the neutron energy calculated from the averaged $^3$He polarization. The  averaged neutron polarization at 0.75~eV was $(36.1\pm0.5)$\%. \\

 \subsection{Nuclear polarization determined by spin-dependent asymmetry}\label{NucPol}
 The $^{139}$La nuclear polarization was determined utilizing the spin-dependent asymmetry at the 2.99~eV $s$-wave resonance of $^{138}$La. The spin-dependent asymmetry at the 2.99~eV resonance, after subtracting of the negative $s$-wave component, was obtained as  
\begin{eqnarray}
\varepsilon_{\rm S}=(5.1\pm0.7)\times 10^{-4}.
\label{expp138}
\end{eqnarray}
The spin-dependent cross section of the 2.99~eV resonance $\sigma^{\rm theo}_{{\rm S}, s}$ can be theoretically described using the resonance parameters listed in Table~\ref{resopara} as 
\begin{eqnarray}
\sigma^{\rm theo}_{{\rm S}, s}=\frac{5\pi}{11k^2}\frac{\Gamma^n_s\Gamma_s}{(E-E_s)^2+(\Gamma_s/2)^2},
\label{theory138}
\end{eqnarray} 
where $E_s$, $\Gamma^n_s$, and $\Gamma_s$ are the resonance energy, neutron width, and total width of the 2.99~eV s-wave resonance, respectively.
The nuclear polarization of $^{138}$La can be determined using Eqs.~\ref{EqAs},~\ref{expp138} and \ref{theory138}, taking into account its natural abundance and the neutron polarization at 2.99~eV, yielding a value of 4.9$\pm$0.7\%.  The target temperature $T_{\rm La}$ was calculated based on a Boltzmann distribution and using the magnetic moment and nuclear spin listed in Table~\ref{nucpara}, resulting in $T_{\rm{La}}=75.7^{+10.2}_{-8.9}~\rm{mK}$, which is consistent with the temperature measured at the cold head of 67~mK. Under the assumption that the spin temperature of $^{139}$La is the same as that of $^{138}$La, the corresponding $^{139}$La nuclear polarization was determined to be $3.9\pm0.5\%$.

\begin{table}[htbp]
\begin{center}
	\begin{tabular}{|c||c|c|c|}
	%\multirow{2}{*}{$r$} \\ 
	%\multicolumn{6}{c||}{published values} \\
	\cline{1-4}
	Isotope&	$I^P$ & 	Abundance & 	$\mu_0$ \\ 
	\hline
	$\rm{^{139}La}$ & $7/2^+$ & $99.91\%$ & $2.78$\\
	$\rm{^{138}La}$ & $5^+$ &0.09\% & $3.71$  \\
	\cline{1-4}
	\end{tabular}
	\caption{
	Parameters of lanthanum isotopes. Nuclear spin and parity $I^P$, natural abundance, and nuclear magnetic moment $\mu_0$ are listed. The unit of the nuclear magnetic moment is nuclear magneton.}
	\label{nucpara}
\end{center}	
\end{table}

\subsection{Spin-dependent cross sections at the resonances}
\begin{figure}[htbp]
	\centering
	\includegraphics[width=0.95\linewidth]{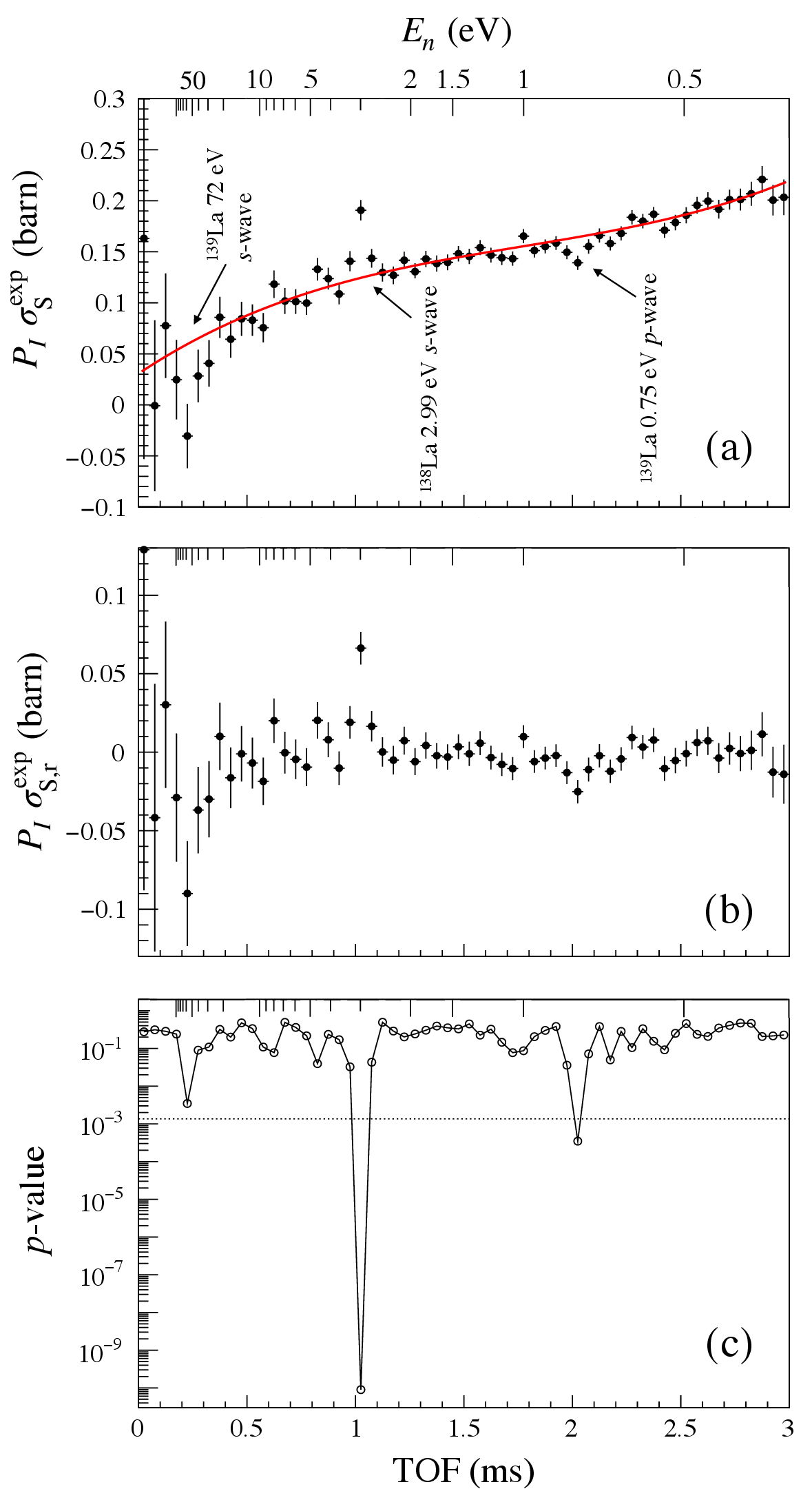}
	\caption[]{	
	(a) TOF dependence of $P_I\sigma^{\rm exp}_{\rm S}$. The curved line is the best fit of the global structure derived from the negative $s$-wave resonance. (b) Resonance component of spin-dependent cross section. (c) $p$-value for $P_I\sigma^{\rm exp}_{{\rm{S}}, r}$. The dotted line shows 99.7\% confidence level.
	}
	\label{Asypvalue}
\end{figure}

The experimental value of the spin-dependent cross section $\sigma^{\rm exp}_{\rm S}$ was obtained from the asymmetry $\varepsilon_{\rm S}$ using Eq.~\ref{EqAs}. The resonance component $\sigma^{\rm {exp}}_{{\rm{S}},r}$ was isolated by fitting the global structure attributed to the negative $s$-wave component with a third order polynomial function. The resonance regions listed in the Table~\ref{resopara} are excluded from the fitting. Figure~\ref{Asypvalue} shows the TOF dependence of $P_I\sigma^{\rm exp}_{\rm S}$ and $P_I \sigma^{\rm exp}_{{\rm{S}},r}$. Note that Fig.~\ref{Asypvalue} was calculated using the areal density of $^{139}$La.  A $p$-value, defined as $p=(1-\rm{C.L.})/2$, where C.L. is the confidence level of the non-zero asymmetry, is also depicted to show the significance of $P_I\sigma^{\rm exp}_{{\rm{S}},r}$ in Fig.~\ref{Asypvalue}. The {\it p}-value indicates the probability to observe a non-zero value of $P_I \sigma^{\rm exp}_{{\rm{S}},r}$ in the hypothesis of no asymmetry. A confidence level of over 99.7\% corresponds to a {\it p}-value less than $1.35 \times 10^{-3}$. The spin-dependence cross section was first observed at the $p$-wave resonance with over 99.7\% C.L. as shown in Fig~\ref{Asypvalue}.\\\\
The spin-dependent cross section in the $p$-wave resonance region of $E_p-3\Gamma_p<E_n<E_p+3\Gamma_p$ after the subtraction of the negative $s$-wave component, defined as $\sigma^{\rm exp}_{{{\rm{S}}, p}}$, is obtained using the nuclear polarization in Section~\ref{NucPol} as 
\begin{eqnarray}
\sigma^{\rm exp}_{{\rm{S}}, p}=-0.26\pm0.08~\rm{barn},
%\sigma^{\rm exp}_{{\rm{S}}, p}=-0.26\pm0.08({\rm{stat.})}^{+0.03}_{-0.00}({\rm{sys.})}~\rm{barn},
\label{expp}
\end{eqnarray}
where $E_p$ and $\Gamma_p$ are the resonance energy and total width of the $p$-wave resonance, shown in Table~\ref{resopara}. Here, the total width is defined as $\Gamma_p=\Gamma_p^\gamma+\Gamma_p^n$. The asymmetry of the spin-dependent cross section relative to the spin-independent cross section of the $p$-wave component was also obtained as
\begin{eqnarray}
A_{\rm S}&=&\frac{\sigma^p_+-\sigma^p_-}{\sigma^p_++\sigma^p_-}=\frac{\sigma^{\rm exp}_{{\rm S},p}}{\sigma^{\rm theo}_{{\rm 0},p}}\nonumber\\
 &=&-0.36\pm0.11.
% &=&0.36\pm0.11({\rm{stat.})}^{+0.04}_{-0.00}({\rm{sys.})}.
\label{asy_sigma}
\end{eqnarray}
The spin-independent cross section $\sigma^{\rm theo}_{{\rm 0},p}$ was theoretically calculated with a Breit-Wigner formula, defined as, 
\begin{eqnarray}
&&\sigma^{\rm {theo}}_{0, p}=\frac{9\pi}{16k^2}\frac{\Gamma^n_p\Gamma_p}{(E-E_p)^2+(\Gamma_p/2)^2}.
\label{theoryp0}
\end{eqnarray}
\\
When using the nuclear polarization calculated from the temperature measured at the cold head, the differences of $\sigma^{\rm exp}_{{\rm{S}}, p}$ and $A_{\rm S}$ from the values in Eq.~\ref{expp} and Eq.~\ref{asy_sigma} were $+0.03~\rm{barn}$ and $+0.04$, respectively. These differences were smaller than the statistical error.
% in Table~\ref{resopara} as 
%\begin{eqnarray}
%&&\int_{E_p\pm 3\Gamma_p}(\sigma'_{{\rm 0},p})dE^{{\rm m}}_n\\
%&=&\int_{E_p\pm 3\Gamma_p}(\frac{4\pi}{32k^2}\frac{g_p\Gamma^n_p\Gamma_p}{(E-E_p)^2+(\Gamma_p/2)^2})dE^{{\rm m}}_n\\
%&=&0.225
%\end{eqnarray}

\section{Analysis}

Under the experimental conditions, the spin-dependent assymetry can be approximated as
	\begin{align}
		\varepsilon_{\rm S} \simeq P_I P_n \rho d \frac{4\pi}{k} {\rm Im}B' ,
	\end{align}
	as described in the Appendix A, where $B'$ is the coefficient in Eq.(10) in Ref.~\cite{Gudkov2020} representing the spin-spin interaction in the forward angle scattering amplitude.
	The following subsections will discuss the implications of the experimental results to the partial neutron width of the $p$-wave resonance and the spins of the $s$-wave resonances.

 \subsection{Determination of partial neutron width using spin-dependent cross section}
The partial neutron width can also be extracted from the angular correlations of $\gamma$-rays emitted from $p$-wave resonances, which arise from interference between $s$- and $p$-wave amplitudes~\cite{okuda18, okuda21, yama20, koga22, endo22}. The advantage of using the spin-dependent cross section is that the neutron partial width can be directly determined without assuming the interference between partial amplitudes and the final state spin after the $\gamma$ decay.\\\\
The spin-dependent cross section at the $p$-wave resonance can be calculated using the explicit theoretical expression of $B'$ as~\cite{Gudkov2020} 
\begin{eqnarray}
\sigma^{\rm {theo}}_{{\rm{S}}, p}&=&\frac{4\pi}{k}{\rm{Im}} B'=\frac{\pi}{16k^2}\frac{\Gamma^n_p\Gamma_p}{(E-E_p)^2+(\Gamma_p/2)^2}\nonumber\\
&&\times \left(-\frac{39}{4}x_{\rm s}^2+\frac{9}{2}\sqrt{\frac{7}{5}}x_{\rm s} y_{\rm s}+\frac{63}{20}y_{\rm s}^2\right),
\label{theoryp}
\end{eqnarray}
where $x_s$ and $y_s$ are ratios of the neutron partial width of the channel spin, defined as 
\begin{eqnarray}
x_{\rm s}&=&\frac{1}{2\sqrt{3}}(-\sqrt{7}x-\sqrt{5}y)\\
y_{\rm s}&=&\frac{1}{2\sqrt{3}}(\sqrt{5}x-\sqrt{7}y).
\label{theryp}
\end{eqnarray}

The neutron partial widths of the neutron total angular momentum $j=1/2$ and $3/2$ components, denoted as $\Gamma^n_{p, j=1/2}$ and $\Gamma^n_{p, j=3/2}$, are expressed by $x$ and $y$ defined as  
\begin{eqnarray}
x^2=\frac{\Gamma^n_{p, j=1/2}}{\Gamma^n_p}, \ \ y^2=\frac{\Gamma^n_{p, j=3/2}}{\Gamma^n_p},
\label{xy}
\end{eqnarray}
where $x$ and $y$ satisfy $x^2+y^2=1$.  The corresponding mixing angle $\phi$ can be defined as 
\begin{eqnarray}
x=\cos\phi, ~y=\sin\phi, 
\label{phi}
\end{eqnarray}
as discussed in Ref~\cite{Gudkov2020}.
The broadening effect by the pulse shape of the neutron beam at 0.75~eV was negligibly small compared with the total width of the $p$-wave resonance and the statistical error, and therefore, the spin-dependent cross section obtained in Eq.~\ref{expp} can be directly compared with the theoretical calculation. By calculating the Breit-Wigner function over the region $E_p-3\Gamma_p<E_n<E_p+3\Gamma_p$ in Eq.~\ref{theoryp}, we obtained the following equation.

\begin{eqnarray}
-0.26\pm0.08=0.079\left(-7x^2-2\sqrt{35}xy+\frac{2}{5}y^2\right)
\label{eqxy}
\end{eqnarray}

Using Eq.~\ref{phi} and Eq.~\ref{eqxy}, we find the solutions for $\phi$ as

\begin{eqnarray}
%\phi=&&(277.4\pm3.1)^\circ, ~(82.5\pm3.1)^\circ,\nonumber\\
%&&(175.3\pm 3.1)^\circ, ~(4.7\pm 3.1)^\circ.
\phi=&&(74\pm4)^\circ,
~(164\pm 4)^\circ, \nonumber\\
&&~(254\pm4)^\circ, ~(344\pm 4)^\circ.
\label{phisolution}
\end{eqnarray}

%\begin{eqnarray}
%(x, y)=&&(-0.130^{+0.054}_{-0.054}, -0.991^{+0.009}_{-0.007}),\nonumber \\
%(&&0.130^{+0.054}_{-0.054}, 0.991^{+0.009}_{-0.007}), \nonumber\\
%&&(-0.996^{+0.007}_{-0.004},~0.093^{+0.054}_{-0.055}), \nonumber\\
%&&(0.996^{+0.004}_{-0.007}, -0.093^{+0.055}_{-0.054})
%\label{xysolution}
%\end{eqnarray}
%Note that the errors were calculated with only the statistical error. 
The corresponding $x$ and $y$ values are also obtained as
\begin{eqnarray}
(x, y)=&&(0.28\pm0.06,~0.96\pm0.02),\nonumber \\
&&(-0.96\pm0.02,~0.28\pm0.06), \nonumber\\
&&(-0.28\pm0.06,~-0.96\pm0.02), \nonumber\\
&&(0.96\pm0.02,~-0.28\pm0.06).
\label{xysolution}
\end{eqnarray}

The visualization of $\phi$ is shown in Fig.~\ref{xy}. 
Equation~\ref{eqxy} is described as the curved line in the $xy$ plane. The intersections of the curved lines and unit circle show the solutions of $\phi$.

\begin{figure}[htbp]
	\centering
	\includegraphics[width=0.80\linewidth]{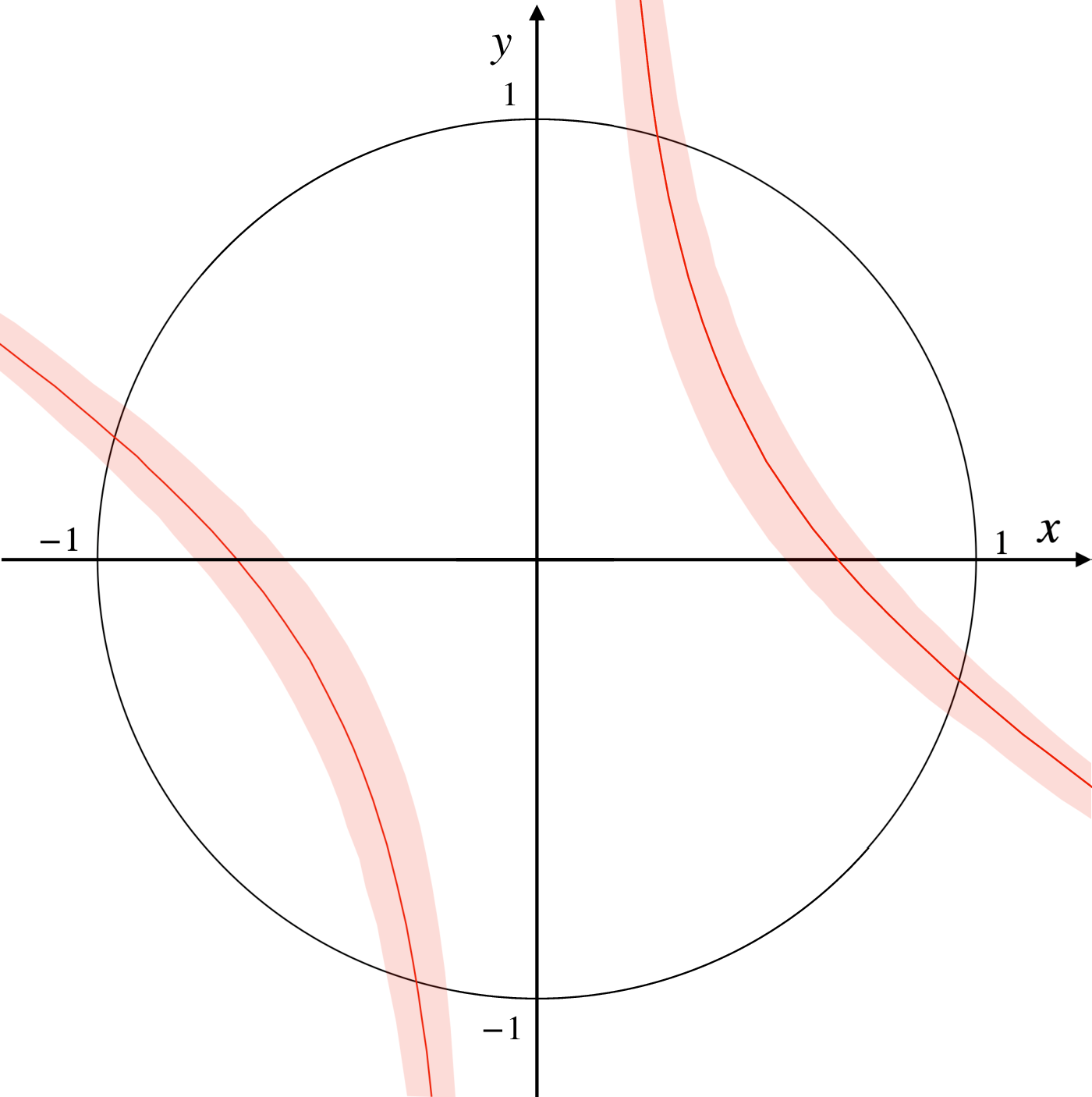}
	\caption[]{
	Visualization of the $\phi$ values on the $xy$ plane. %The curved lines and shaded area shows the Eq.~\ref{eqxy} and its 1$\sigma$ region.  
	The curved lines, shaded areas, and dashed lines show Eq.~\ref{eqxy} and its 1$\sigma$ region of the statistic error.  
	}
	\label{xy}
\end{figure}
The above analysis was also performed using the resonance parameters reported by other groups in Appendix~\ref{AppendixAna}. The differences in the analysis results that arose from differences in the resonance parameters were within the statistical error. We confirmed that these differences stemming from the resonance parameters do not affect the conclusions of this paper.
 
\subsection{Spin of $s$-wave resonances}
For the $s$-wave resonances, the spin $J$ can be directly determined from the asymmetry. The positive (negative) sign of the asymmetry indicates that neutrons with parallel (anti-parallel) spin are likely to be absorbed by nuclei. The sign of the asymmetry in Fig.~\ref{Asypvalue} (a) and (b) implies: $J=4$ for the negative $s$-wave resonance of $^{139}$La, whose spin is $7/2$; $J=11/2$ for the 2.99~eV $s$-wave resonance of $^{138}$La, whose spin is 5; and  $J=3$ for the 72.3~eV $s$-wave resonance of $^{139}$La. The spins of the $s$-wave resonances determined in this experiment are consistent with the reference values in Table~\ref{resopara}.

\section{Conclusion}
We observed the spin-dependent cross section at the 0.75~eV $p$-wave of $^{139}$La+$n$ using a polarized lanthanum target and a polarized pulsed neutron beam.  The partial neutron width of the $p$-wave resonance was determined. In a separate paper, these results will be compared with other experimental results of ($n$,$\gamma$) reactions~\cite{okuda18, okuda21, yama20, koga22, endo22, okuda2023} in terms of the $s$-$p$ mixing model and will be used for improving a quantitative understanding the symmetry violation enhancement mechanism.

\begin{acknowledgments}
The authors would like to thank the staff of beamline 22 for the maintenance, the low temperature sample environment team for the operation of the superconducting magnet and the dilution refrigerator, and MLF and J-PARC for operating the accelerators and the neutron production target. T. Okudaira would like to especially thank S.~Ohira-Kawamura and M. Matsuura for their assistance designing the La holder. The neutron scattering experiment was approved by the Neutron Scattering Program Advisory Committee of IMSS and KEK (Proposals Nos. 2018S12). The neutron experiment at the Materials and Life Science Experimental Facility of the J-PARC was performed under a user program (Proposal No. 2022A0101). This work was supported by JSPS KAKENHI Grant Nos. 20K14495, 23K13122, JST SPRING Grant No. JPMJSP2125, and the US National Science Foundation PHY-1913789 and PHY-2209481. R. Nakabe acknowledges support from the Interdisciplinary Frontier Next-Generation Researcher Program of the Tokai Higher Education and Research System. W. M. Snow acknowledges support from the Indiana University Center for Spacetime Symmetries. J.~G.~Otero~Munoz acknowledges support from the National GEM consortium. V.~Gudkov acknowledges support from the U.S. Department of Energy Office of Science, Office of Nuclear Physics program under Award No. DE-SC0020687.
\end{acknowledgments}

\appendix

\section{Neutron spin behavior in the polarized target and its approximation}
We employ the optical description of the neutron spin behavior in polarized target material as described in Ref.~\cite{Kabir1988, STODOLSKY19865} to describe the measured asymmetry  $\varepsilon_{\rm S}$ as
\begin{align}
\varepsilon_{\rm S} &= P_n\frac
	{{\rm Tr}(\mathfrak{S}^{\dag}\sigma_x\mathfrak{S})}
	{{\rm Tr}(\mathfrak{S}^{\dag}\mathfrak{S})}
%\nonumber\\
=P_n\frac{2{\rm Re}A^{\ast}B+{2\rm Im}C^{\ast}D}{\abs{A}^2+\abs{B}^2+\abs{C}^2+\abs{D}^2} .
\label{eq:eps1}
\end{align}

The coefficients $A$, $B$, $C$, and $D$ are related to the forward angle scattering amplitude given in Ref.~\cite{Gudkov2020} as
\begin{align}
A &= e^{i\alpha}\cos\beta
	,\quad
%	\nonumber\\
B = ie^{i\alpha}\frac{\sin\beta}{\beta} \beta_x
	\nonumber\\
C &= ie^{i\alpha}\frac{\sin\beta}{\beta} \beta_z
	,\quad
%	\nonumber\\
D = ie^{i\alpha}\frac{\sin\beta}{\beta} \beta_y
\end{align}
where
\begin{align}
\frac{\alpha}{Z} &= A' + P_1H'(\bm{k}\cdot\bm{I})+P_2E'((\bm{k}\cdot\bm{I})^2-\frac13)
	\nonumber\\
\frac{\beta_x}{Z} &=P_1B'+\frac{\mu_{\rm n}m_{\rm n}}{2\pi\hbar^2\rho}B_{\rm ext}+P_2F'(\bm{k}\cdot\bm{I})+P_3\frac{B_3'}{3}((\bm{k}\cdot\bm{I})^2-1)
	\nonumber\\
\frac{\beta_y}{Z} &=P_1D'+P_2G'(\bm{k}\cdot\bm{I})
	\nonumber\\
\frac{\beta_z}{Z} &=C'+P_1K'(\bm{k}\cdot\bm{I})-P_2\frac{F'}{3}+P_3\frac{2B_3'}{3}(\bm{k}\cdot\bm{I})
	\nonumber\\
\beta^2 &= \beta_x^2 + \beta_y^2 + \beta_z^2
,\quad
%	\nonumber\\
Z = \frac{2\pi\rho d}{k} .
\end{align}

Here, $B_{\rm{ext}}$, $\mu_n$, and $m_n$ denote the external magnetic field, neutron magnetic moment, and mass, respectively. $\bm{
k}$ and $\bm{I}$ are unit vectors parallel to the neutron momentum and the nuclear spin. The $P_1$, $P_2$ and $P_3$ represent the target nuclear polarization of 1st-rank (vector), 2nd-rank, and 3rd-rank spherical tensors, respectively, and amounts $P_1=P_I= 3.9 \pm 0.5\%$, $P_2 = 0.10^{+0.03}_{-0.02}\%$, $P_3 = (2.1\pm1.0) \times 10^{-3}\%$ in the present experimental conditions.
Non-zero values of the $(\bm{k}\cdot\bm{I})$ originate from the beam divergence up to the maximum value of $2 \times 10^{-3}$.\\
The valuables $A'$-$G'$ are the coefficients of correlation terms in the forward scattering amplitude for polarized $^{139}$La nuclei and polarized neutrons defined as ~\cite{Gudkov2020}:
\begin{equation}
 \begin{split}
  f=&A'+P_1H'({\bm{k}}\cdot{\bm{I}})+P_2E'\left({(\bm{k}}\cdot{\bm{I}})^2-\frac{1}{3}\right)\\
   &+(\bm{\sigma}\cdot{\bm{I}})\left(
   P_1B'+P_2F'({\bm{k}\cdot{\bm{I}}})+P_3\frac{B_3'}{3}\left(({\bm{k}\cdot{\bm{I}}})^2-1\right)
   \right)\\
  &+(\bm{\sigma}\cdot{\bm{k}})\left(
   C'+P_1K'({\bm{k}}\cdot{\bm{I}})-P_2\frac{F'}{3}+P_3\frac{2B_3'}{3}({\bm{k}\cdot{\bm{I}}})
   \right)\\
  &+\bm{\sigma}\cdot({\bm{k}\times {\bm{I}}})\left(
   P_1D'+P_2G'({\bm{k}}\cdot{\bm{I}}).
   \right)
   \label{fdash}
 \end{split}
\end{equation}

The magnitude of coefficients $H'$, $E'$, $F'$, $B_3'$, $K'$ and $G'$ are of the same order or less than $A'$ and $B'$ on the basis of the explicit expressions in Eq.~(28)-(37) of Ref.~\cite{Gudkov2020}.
The magnitudes of P-odd, T-even term $C'$ and P-odd, T-odd term $D'$ are smaller than $A'$ and $B'$ by more than two orders of magnitudes.
Consequently, the value of $\beta$ can be approximated as $\beta\simeq\beta_x$ and we obtain
\begin{align}
\varepsilon_{\rm S}
%&\simeq -P_n\frac{4\pi\rho d}{k}\sinh({\rm Im}\beta_x)\cosh({\rm Im}\beta_x) .
&\simeq P_n\tanh(2{\rm Im}\beta_x).
\end{align}
The numerical value of ${\rm Im}\beta_x$ is about $10^{-3}$, which leads to
\begin{align}
\varepsilon_{\rm S}
\simeq P_I P_n\rho d\frac{4\pi}{k}{\rm Im}B' .
\end{align}
\\
\section{Analysis using resonance parameters reported in other references}
\label{AppendixAna}
For the 0.75~eV $p$-wave resonance, the measurements using neutron transmission or ($n$, $\gamma$) reaction have been reported by several groups listed in Table~\ref{refpara}~\cite{JENDL-La-28, Mughabghab-La-Capture-05, Mughabghab-La-Capture-15, Exfor-La-Capture-16}. The details of each measurement of the resonance parameters are summarized in Ref.~\cite{EndoReso2023}.

%Recently, Endo {\it{et al}}., measured new resonance parameters using the (n, $\gamma$) reaction and neutron transmission with the high-intensity neutron beam at J-PARC~\cite{EndoReso2023}. \\

\begin{table*}[htbp]
\centering
%\begin{ruledtabular}
 \begin{tabular}{|cc|c|c|c|}
 \hline
 & Reference & $E_0$~[eV] & $\Gamma_\gamma$~[meV] & $g\Gamma_n$~[meV] \\ 
 \hline
   & Endo {\it{et al.}} (2023)~\cite{EndoReso2023} & $0.750\pm0.001$ & $41.6\pm0.9$ & $(3.67\pm0.05)\times10^{-5}$ \\ 
  &Terlizzi {\it{et al.}} (2007)~\cite{JENDL-La-28}      & $0.758\pm0.001$     & $40.1\pm1.9$     & $(5.6\pm0.5)\times10^{-5}$    \\
  & Alfimenkov {\it{et al.}} (1983)~\cite{Mughabghab-La-Capture-05}  & $0.75\pm0.01$ & $45\pm5$ & $(3.6\pm0.3)\times 10^{-5}$ \\
  &Shwe {\it{et al.}}  (1967)~\cite{Mughabghab-La-Capture-15}     & $0.734\pm0.005$      & $40\pm5$     & $(3.67\pm0.22)\times10^{-5}$    \\
  &Harvey {\it{et al.}} (1959)~\cite{Exfor-La-Capture-16}    & $0.752\pm0.011$      & $55\pm10$     & $(4\pm1)\times10^{-5}$    \\
   \hline
 \end{tabular}
 \label{refpara}
 \caption{\label{refpara} Resonance parameters of $^{139}$La $p$-wave resonance reported in each reference.}
%\end{ruledtabular}
\end{table*}

\begin{table*}[htbp]
\begin{center}
	\begin{tabular}{|c||c|c|c|}
	%\multirow{2}{*}{$r$} \\ 
	%\multicolumn{6}{c||}{published values} \\
	\cline{1-4}
	\begin{tabular}{c}Reference for \\resonance parameters \end{tabular}&$\sigma^{\rm {exp}}_{{\rm{S}}, p}$~[barn] & $A_{\rm S}$& 	$\phi~[\rm{degree}]$ \\ 
	\hline
	Endo {\it{et al.}} & $-0.26 \pm 0.07$ & $-0.36 \pm 0.10$ & $~74\pm4,~164 \pm 4,~254\pm 4,~344\pm4$\\
	Terlizzi {\it{et al.}} &  $-0.26 \pm 0.07$  &$-0.24 \pm 0.06$ & $~79\pm2,~159 \pm 2,~259\pm 2,~339\pm2$  \\
	Alfimenkov {\it{et al.}}& $-0.23 \pm 0.06$  &$-0.36 \pm 0.10$ & $~74\pm4,~164 \pm 4,~254\pm 4, ~344\pm4$\\
	Shwe {\it{et al.}}& $-0.26 \pm 0.07$  &$-0.35 \pm 0.09$ & $~75\pm4,~163 \pm 4,~255\pm 4,~343\pm4$\\
	Harvey {\it{et al.}}& $-0.19 \pm 0.06$  &$-0.33 \pm 0.10$ & $~76\pm4,~163 \pm 4,~256\pm 4, ~343\pm4$\\
	\cline{1-4}
	\end{tabular}
	\caption{
	Analysis results using the resonance parameters in each reference.}
	\label{Ana_refResoPara}
\end{center}	
\end{table*}
 
\begin{figure}[htbp]
	\centering
	\includegraphics[width=0.80\linewidth]{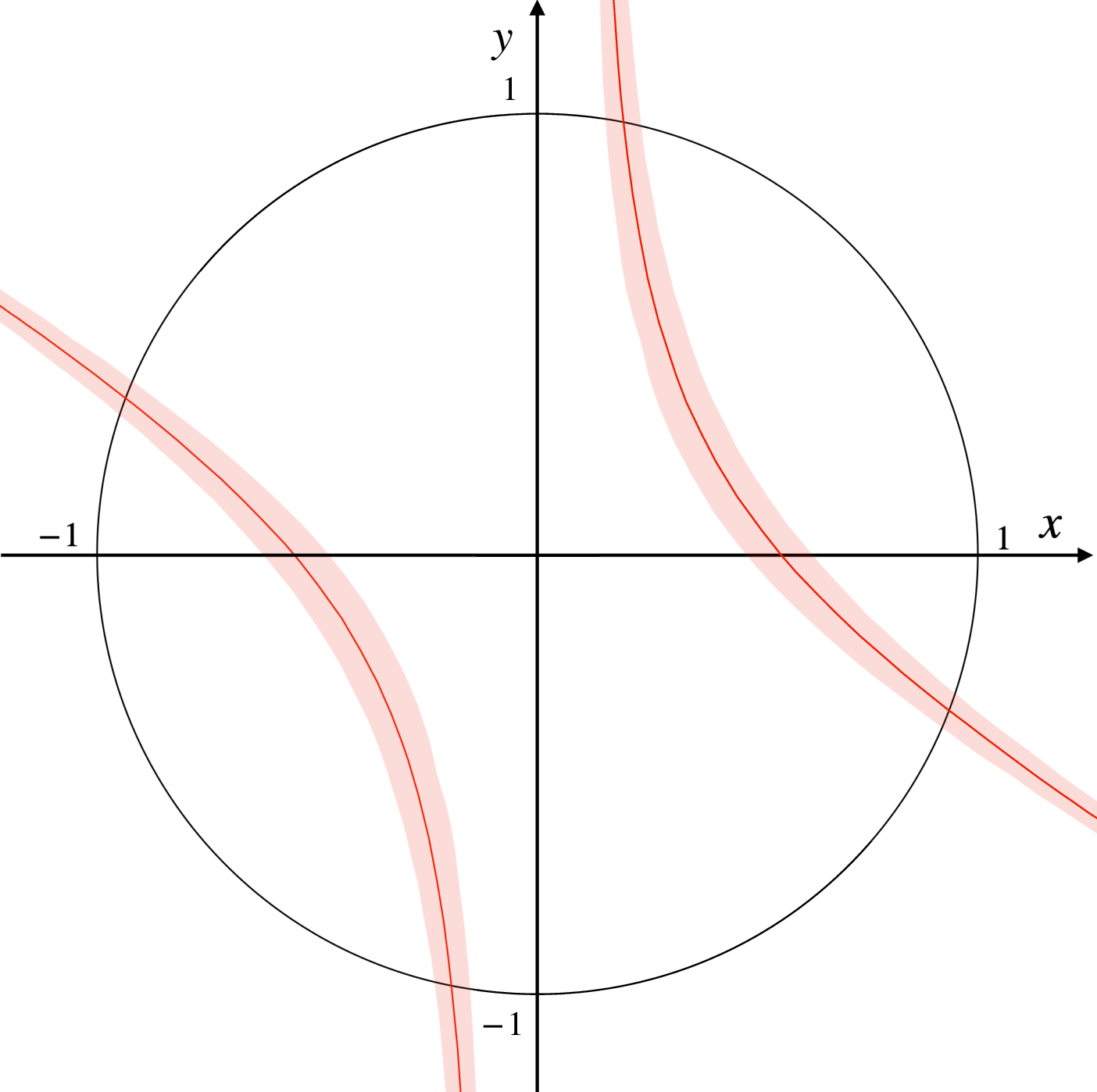}
	\caption[]{
	Visualization of the $\phi$ values obtained using Terilzzi {\it{et al.}}'s resonance parameters on the $xy$ plane. The curved lines and shaded area shows Eq.~\ref{eqxy} and its 1$\sigma$ region of the statistic error.  
	}
	\label{xy_Terlizzi}
\end{figure}

Tables~\ref{Ana_refResoPara} shows $\sigma^{\rm exp}_{{\rm S}, p}$, $A_{\rm S}$, and $\phi$ values obtained using resonance parameters reported by each group. 
The central values for $A_{\rm S}$ show agreement within an accuracy of 10\% or less with the exception of that based on resonance parameters reported by Terilzzi {\it{et al.}}. The central value of $A_{\rm S}$ using this resonance parameters exhibit a difference of around 30\%, which is attributed to Terilzzi {\it{et al.}}'s $g\Gamma_n$ being reported as approximately 30\% larger than in other references. Consequently, the $\phi$ values obtained using resonance parameters reported by Terilzzi {\it{et al.}} show difference compared to the analysis using other resonance parameters, as illustrated in Fig.~\ref{xy_Terlizzi}. However, these differences remain consistent within the statistical errors obtained in the present experiment.

\bibliography{ImB}
\end{document}